\def\vsp#1{}
\documentclass{elsart}
\usepackage{graphicx}
\usepackage{amsmath}
\usepackage{amsfonts}
\usepackage{amssymb}

\begin{document}
\begin{frontmatter}
\title{EQSE Diagonalization of the Hubbard Model}
\author{Nathan Salwen}
\address{Dept. of Physics, Harvard University, Cambridge MA 02138,\\
salwen@physics.harvard.edu}
\begin{abstract}
  The application of enhanced quasi-sparse eigenvector methods (EQSE)
  to the Hubbard model is attempted. The ground state energy for the
  4x4 Hubbard model is calculated with a relatively small set of basis
  vectors.  The results agree to high precision with the exact answer.
  For the 8x8 case, exact answers are not available but a simple first
  order correction to the quasi-sparse eigenvector (QSE) result is
  presented.
\end{abstract}
\end{frontmatter}

\section{Introduction}

The enhanced quasi-sparse eigenvector (EQSE) method of solving quantum
field theory Hamiltonians is the combination of quasi-sparse
eigenvector (QSE) method \cite{qse} with a stochastic calculation for
the contribution of the remaining basis states \cite{eqse}.  The
Hubbard model was chosen as a laboratory for testing this method for
several reasons. First, we believed the approach could yield results.
The basis vectors can be specified in a few words of data and the
Hamiltonian is sparse is momentum space. On the other hand, the ground
state of the Hubbard model is known for its inclusion of a
extraordinary number of Fock states \cite{and} so the model presents a
non-trivial challenge to the quasi-sparse approach. Finally, the
Hubbard model is thought to be a physically relevant model for
superconductivity \cite{and}. While the description of the model is
simple, solutions have been difficult and any promising new approach
is worthwhile.

We work on a 2-dimensional spatial lattice with the Hubbard Hamiltonian
\[
H=-t\sum _{\footnotesize \begin{array}{c}
\sigma =\uparrow ,\downarrow \\
<i,j>\text {nearest}
\end{array}}(c_{i\sigma }^{\dagger }c_{j\sigma }+c_{j\sigma }^{\dagger }c_{i\sigma })+U\sum (c_{i\uparrow }^{\dagger }c_{i\uparrow }c_{i\downarrow }^{\dagger }c_{i\downarrow })\]

There are 2 species of electron so there are 4 possible states for each lattice
site. Thus the 8x8 Hubbard model has \( 4^{64} \) dimensions. Even after using
particle conservation to partition the space there are more than \( 10^{32} \)
basis vectors. In this large space the Hamiltonian is clearly sparse But the
equality of the off-diagonal elements contributes to an extraordinary number
of Fock states in the ground.

It is thought that the D-wave correlator \( C_{d_{x^{2}-y^{2}}}(r) \) is an
important indicator of superconductivity \cite{huss}.

\section{Hamiltonian Momentum Lattice Formulation}

After making space periodic we use the Fock states as our basis set. The Hamiltonian
conserves momentum which helps limit the number of relevant basis states. 
\begin{align*}
H_{\text {kin}}=
-2t\sum (\cos \frac{\pi n}{L}+\cos \frac{\pi  n}{L})
a_{np,\sigma }^{\dagger }a_{np,\sigma } \\
V=\frac{U}{4L^{2}}
\sum _{
  \footnotesize
  \begin{array}{c}
    k-l+m-n=0\\
    p-q+r-s=0
  \end{array}
  }
a_{kp}^{\uparrow \dagger }a^{\uparrow }_{lq}a_{mr}^{\downarrow \dagger }a^{\downarrow }_{ns}  
\end{align*}
There is a 16 member symmetry group generated by reflections in the \( x \)
and \( y \) planes, \( x\leftrightarrow y \) and\( \downarrow \leftrightarrow \uparrow  \).
For the purpose of finding the ground state, we use only symmetrized basis states.

The first step in the calcualation consists of picking a basis set of size \( N \)
and diagonalizing its submatrix of the Hamiltonian using the Lanczos
method \cite{arpack}. 
The \( \frac{N}{5} \)basis vectors which least contribute to the ground are
then discarded, replacements are chosen and the Hamiltonian is again diagonalized.
When the ground energy \( E_{0} \) obtained in this manner converges
the QSE step is complete.

The EQSE is a first order correction to this result.  We calculate it
stochastically. Let $C$ be a set of basis vectors for the complement
of the $N$-dimensional subspace.  Then, choosing representative basis
vectors \( v \in C \) with probability \( P(v) \) .

\begin{align*}
E=E_{0}+\text {Average}\frac{FO(v)}{P(v)}, \qquad \qquad
FO(v)=-\frac{\langle 0|H|v\rangle \langle v|H|0\rangle }{\lambda _{v}}
\end{align*}
where \( |0\rangle  \) is the ground state of the \( N \) dimensional subspace. 
The expectation is thus
\[
E=E_{0}+\sum_{v\in C} P(v) \frac{FO(v)}{P(v)}.\]

\section{Results}
Results were obtained for the ground state energy, wavefunction and
$d$-wave correlator.  The computing time was about 2 days on a 350Mhz
Pentium II. Where available, we compare with Husslein et al
\cite{huss} results labeled Exact, Projector Quantum Monte-Carlo
(PQMC), and Stochastic Diagonalization (SD) (which uses a different method
of choosing the subspace than QSE).

{\centering \begin{tabular}{|c|c|c|c|c|c|c|}
\multicolumn{7}c{Ground State Energy 4x4 Hubbard Model($\frac{5}{16}$ occupied) }\\
\hline 
Coupling &
States&
QSE&
EQSE&
Exact&
SD&
PQMC\\
\hline 
\hline 
&50 & -.47471 & -.50127(5)& && \\
U=2&100 & -.47967 & -.50147(5)&
-.50194&
-.501&
-.49\\
&500 & -.49454 & -.50181(3) & & &\\
&1000 & -.50062 & -.50198 & &&\\
\hline
& 50 & -1.5707 & -1.80635(5) & && \\
U=4&100 & -1.6203 & -1.8113(4) & -1.8309 & -1.829 & -1.8(2) \\
&500 & -1.7476 & -1.8242(3) & & & \\
&1000 & -1.8003 & -1.8302 & && \\
\hline
&50 & -2.2450 & -2.6663(8) & && \\
U=5&100 & -2.3322 & -2.6724(7)&
-2.7245&
-2.723&
-2.9(3) \\
&500 & -2.5578 & -2.7073(4) & && \\
&1000 & -2.6512 & -2.7208(2)  & &&\\
&2000 & -2.685 & -2.7231  & &&\\
\hline 
&50 & -2.963 & -3.615 & && \\
U=6&100 & -3.103 & -3.635 & && \\
&1000 & -3.452 & -3.697 & && \\
&2000 & -3.595 & -3.723 & && \\
\hline 
\end{tabular}\par}
\vsp{0.3cm}

\vsp{0.3cm}
{\centering \begin{tabular}{|c|c|c|c|}
\multicolumn{4}c{Ground Energy 8x8 Hubbard Model($\frac{25}{64}$ occupied)}\\
\hline 
Coupling &
States&
QSE&
EQSE\\
\hline 
\hline 
&50 & -.281 & -2.58(2)\\
&100 & -.443 & -2.49(2)\\
U=2&500 & -1.221 & -2.40(2)\\
&1000 & -1.751 & -2.406(8)\\
&2000 & -1.956 & -2.423(9)\\
&4000 & -1.958 & -2.427 \\
\hline
& 50 & -.811 & -9.18(6)  \\
&100 & -1.386 & -8.48(6)\\
U=4&500 & -3.449 & -7.58(5) \\
&1000 & -4.798 & -7.59(3) \\
&2000 & -5.374 & -7.620(4) \\
&4000 & -5.387 & -7.621(5) \\
\hline
&50 & -1.222 & -13.33(8) \\
&100 & -1.88 & -12.33(8) \\
U=5&500 & -4.65 & -10.65(6) \\
&1000 & -6.44 & -10.65(3)\\
&2000 & -7.225 & -10.671(5)\\
&4000 & -7.236 & -10.662(8)\\
\hline 
&50 & -1.6 & -18.2 \\
&100 & -2.4 & -16.6 \\
U=6&500 & -5.9 & -13.93(8) \\
&1000 & -8.15 & -13.78(4) \\
&2000 & -9.110 & -13.888(7) \\
&4000 & -9.113 & -13.880(8) \\
\hline 
\end{tabular}\par}
\vsp{0.3cm}

The $d_{x^2-y^2}$ correlator was also obtained using the QSE
algorithm.  The 4x4 result again matched that of Husslein et al
\cite{huss}. The EQSE calculation has not been completed and we
therefore omit the data.

\section{Conclusion}

As we can see, the ground state for the 4x4 Hubbard model can be well
described with about 1000 symmetrized states and the 50 state results
yield remarkable accuracy when the first order correction is included.
In the 8x8 case it is clear that even 4000 states are not sufficient
to describe the ground state.  The precision of the first-order
values will be determined with the completion of higher order
calculations. 

Further advances will come in the refinement of the enhancement
technique.  Better importance sampling will speed convergence of
second and higher orders contributions.  Other extensions will be
calculation of excited states of the Hamiltonian, correlation
functions, binding energies and other quantities of interest.

\paragraph*{Acknowledgments}

The author thanks the organizers of the Light-Cone Meeting at
Heidelberg, Nikolay Prokofyev and collaborators on the original papers
cited here.

\end{document}